\documentclass[10pt,a4paper]{article}
\usepackage{graphicx}

\begin{document}
\title{Solvent-induced symmetry breaking and second order phase transitions}
\author{ F. S. Zhang$^{1,2}$ and R. M. Lynden-Bell,$^{1}$
\thanks{e-mail
address:rmlb@cam.ac.uk; current address: University Chemical Laboratory, Lensfield Road, Cambridge CB2 1EW, UK}\\ \\
$^1$Atomistic Simulation Group, School of Mathematics and Physics, \\
Queen's University, Belfast BT7 1NN, UK\\
$^2$Institute of Modern Physics, Chinese Academy of Sciences,\\
                         P. O. Box 31, 730000 Lanzhou, China }

\date{\today}
\maketitle

\begin{abstract} The triiodide ion is an example of a system where
  symmetry-breaking may be induced by a solvent. The Landau free
  energy is expected to have a similar form to that for the mean field Ising
  model, but with solvent strength rather than temperature as the
  control parameter determining whether there is symmetry
  breaking. In order to examine the possibility of critical phenomena
  we have studied the properties of the ion in solvents based on a
  model for water with charges scaled by a factor $\lambda$. As
  $\lambda$ is increased the system changes from one with no symmetry
  breaking to one with strong symmetry breaking. The behavior of
  various quantities, including the Shannon entropy, as a function of
  $\lambda$ show
only weak maxima near the critical value of
  $\lambda=\lambda_c$. We examine the behavior of a simple model and
  show that divergences would only be expected in the limit of low
  temperatures, and the essential difference between the
  solvent-induced symmetry breaking and the mean field Ising model is
  that in the latter the observed quantity is an average over many
  spins, while in the former observations are made on individual
  molecules.\end{abstract}

\section{Introduction} In an earlier paper we showed that symmetry
  breaking could be
induced in the triiodide ion by varying the
  solvent \cite{zlbprl}.
Experiments and simulations
  \cite{jmay,lbkos,mayth,lbmarg,lbmargcn}
suggest that protic
  solvents which can form hydrogen bonds with a
negative ion cause
  symmetry breaking of the ion, so that the charge
becomes
  concentrated at one end of the ion and the corresponding
  bond
elongates.  We suggested that one could draw an analogy
  between the mean
field Ising model with free
  energy
\begin{equation}
F=a (T-T_0) \eta^2 + C \eta^4
  \label{eq:isingfe}
\end{equation}
and  solvent induced symmetry
  breaking with
\begin{equation}
F=a (R_0-R) \eta^2 + C
  \eta^4. \label{eq:solfe}
\end{equation}
In these expressions $F$
  is the Landau free energy per molecule
(or per spin), $\eta$ is an
  order parameter and $a$ and $C$ are
coefficients \cite{landau}. The
  Ising model has a critical point when the
temperature $T$ reaches
  $T_0$; below this temperature the symmetry
is broken to give
  domains with non-zero order parameters, while
above this
  temperature the order parameter is equal to zero.
Various critical
  phenomena, such as vanishing of the
susceptibility and diverging
  fluctuations in the order parameter,
are observed as the critical
  point is approached. As these
mean-field critical phenomena result
  from the form of the equation
for the free energy, it is
  interesting to see whether there are
corresponding phenomena in the
  solvent-induced symmetry breaking.
In this paper we describe an
  investigation of this point. In order
to vary the solvent strength
  in a systematic way we used a series
of modified waters as
  solvents. The models are based on the
standard spc/e model with a
  Lennard-Jones center on the oxygen
atom and charges on the atomic
  sites. The charges are scaled by a
factor $\lambda$ which varies
  from unity, giving the standard
spc/e model which we know causes
  strong symmetry-breaking, to
zero, giving a pure Lennard-Jones
  solvent which our earlier work
shows does not cause symmetry
  breaking. Thus varying $\lambda$
provides a method of tuning the
  solvent strength $R$ through the
critical value $R_0$.

\section{Theory and computational details}
\subsection{Valence bond
  model for I$^-_3$}
The model for triiodide ion is the same as used
  in our previous
work \cite{zlbprl,lbmarg,lbmargcn,fs1,fs2}. The
  electronic structure of
the ion is described using a semi-empirical
  valence bond model
based on diatomics in molecules method
  \cite{dim} with additional
terms due to the fact that the species
  is charged \cite{dimq}.
Provided the ion is constrained to be
  linear, its ground state is
described by a $3 \times 3$ Hamiltonian
  matrix, whose matrix
elements depend on the instantaneous values of
  the bond lengths
and the instantaneous external electrostatic
  potential due to the
solvent. Full details are given in the
  appendix to reference
\cite{fs2}. The solvent molecules are rigid
  three-site models
  based on spc/e
water \cite{spce},  with partial charges on all three
  atomic sites
and Lennard-Jones interactions on the oxygen site. The
  total
energy of the system can be written
  as
\begin{equation}
E(\{r_i\})=\sum_{\alpha \beta} c_{0
  \alpha}c_{0 \beta} H_{\alpha \beta} +  \sum_{ij} {q_jq_k \over 4
  \pi \epsilon_0 r_{jk}} + \sum_{jk} V^{LJ}_{jk}(r_{jk})
  + \sum_{jm} V^{LJ}_{jI}(r_{im}).
\label{eq:energy} \end{equation}
where the first term is the quantum mechanical energy of the
  ground
state, with $H_{\alpha \beta}$ being the Hamiltonian matrix
  element between
basis states $\alpha$ and $\beta$ and $c_{0
  \alpha}$ the
coefficient of basis state $\alpha$ in the ground
  state. The second
and third terms are  sums of the electrostatic
  and Lennard-Jones
interactions over all pairs of solvent sites $j$
  and $k$, and the last term is the sum of Lennard-Jones interactions
  between solvent
sites and iodine atomic sites $m$.

Quantum
  mechanical forces on both iodine and solvent sites were calculated
  using the Hellman-Feynman theorem, while the forces arising from the
  three classical terms in equation (\ref{eq:energy}) were calculated
  in the usual way within the molecular dynamics program. The ion was
  constrained to be linear throughout.

\subsection{Modified water models}
The solvent molecules were based on the standard spc/e model for a
  water
molecule \cite{spce} with a Lennard-Jones center on the
  oxygen and
charges on the atomic sites. In spc/e water hydrogen
  bonds and all
other orientational correlations are due solely to
  electrostatic
interactions between molecules. Thus the
  hydrogen-bond strength can be
controlled by scaling the charges. In
  this study,  nine solvent models
were used with the charges were
  scaled by a factor
$\lambda$ which varied from 1.125, giving a
  super strong water, through 1,
giving the standard spc/e model, to
  zero, giving a pure Lennard-Jones
solvent.

The Lennard-Jones
  potential between the iodine and oxygen sites
was the same as used
  in our previous work with water
\cite{zlbprl,lbkos,lbmarg}. The
  Lennard-Jones parameters and the partial
charges for sites are
  given in Table \ref{tab:params}.
\subsection{Simulation
  details}
Molecular dynamics simulations were carried out using a
  version of the
dlpoly program \cite{dlpoly} which was modified to
  include the
construction and diagonalisation of the  Hamiltonian
  matrix and the
calculation of the Hellman-Feynman forces. The
  simulation cell contained
one triiodide ion and 509 modified water
  molecules in a cubic box with an
edge-length of 24.8\AA. The
  density was the same for all simulations.
Simulations were carried
  out at 300K with a time step of 1 fs. For each
value of the charge
  scaling constant,
$\lambda$, the system was thoroughly equilibrated
  before
collecting data for 1 ns.  Our earlier work showed that two
  order
parameters were necessary to describe the symmetry
  breaking. The order
parameters chosen  are the molecular dipole
  moment
$\mu$ (relative to the center of mass) and the normal
  coordinate for the
antisymmetric stretch
$\zeta = (
  b_{12}-b_{23})/\sqrt{6}$. The former measures the extent
  of
electronic distortion, while the latter measures
  the
geometrical distortion. The instantaneous values of these
  parameters were
determined at each time step, and averages,
  mean square fluctuations and
probability histograms were
  constructed.

\subsection{Shannon's information entropy}

From the two-dimensional
  histograms for the probability distributions of the two order
  parameters, the Shannon information entropy  function
  $H(\zeta,\mu)$ was constructed for each
  bin 
\begin{equation}
H(\zeta,\mu)=-p(\zeta,\mu)\ln
  p(\zeta,\mu),
\end{equation}
where $p(\zeta,\mu)$ is the
  probability of being in
that bin, with $\sum
  \sum
p(\zeta,\mu)=1$. The Shannon entropy of the
  system \cite{denbigh} is then
\begin{equation}
h_S=\sum_\zeta
  \sum_\mu H(\zeta,\mu).
\end{equation}
We show in the appendix that
  the limit of this sum as the bin sizes tend to zero is a property of the
  system, rather than of our information about it, and is the entropy
  associated with the spread of order parameter values. In our
  calculations the bin sizes
  used were
$\delta \mu = 3D$ and $\delta \zeta= 0.04\AA $. 
  The values of the
bin sizes affect the zero of entropy, but, as they were
  kept constant for all the
simulations, entropy differences
  between runs are real although
absolute values are
  arbitrary.

\section{Results}

\begin{figure}[h!]
\includegraphics[width=12cm]{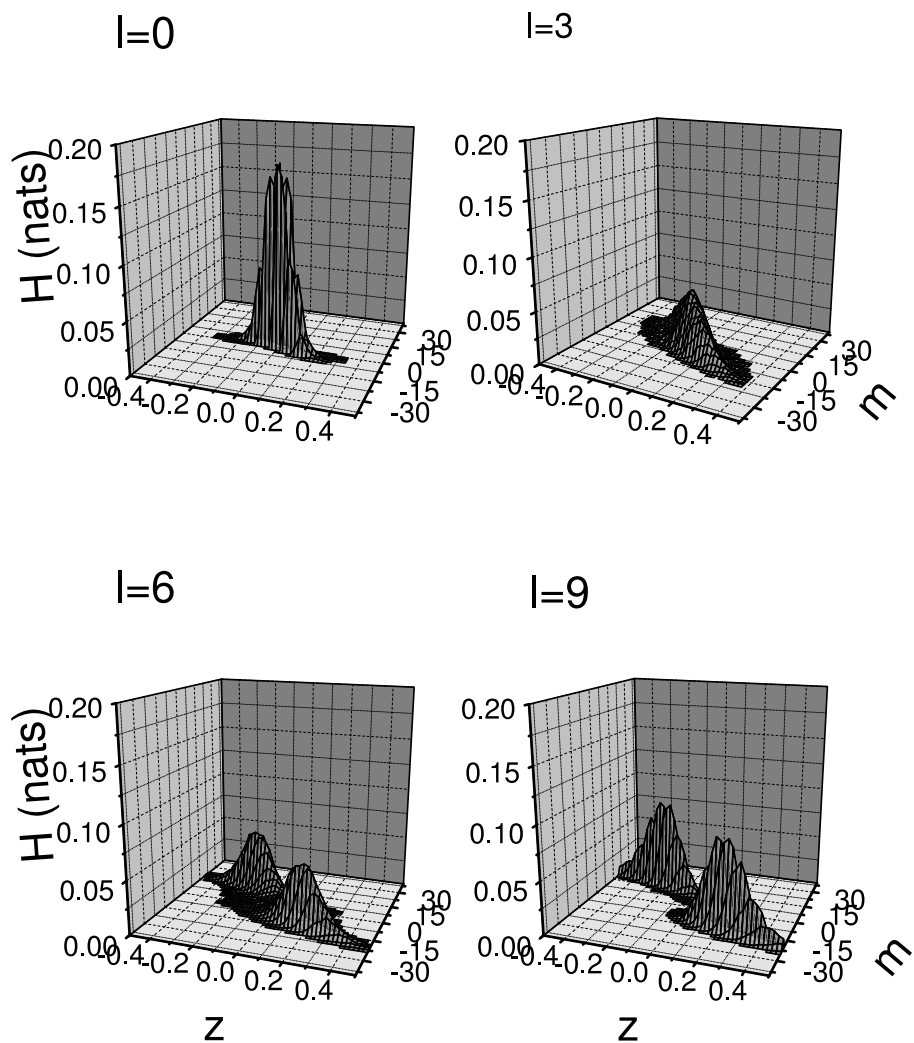}
\caption{Three-dimensional
  plots of
Shannon's information entropy function $H(\zeta,\mu)$ of
  I$^-_3$ at 300 K as a function of the antisymmetric vibrational
  normal coordinate
$\zeta$  and the dipole moment $\mu$ for
  different modified water
solvents. The distributions correspond to
  charge scaling factors
$\lambda=0, 0.375, 0.75, 1.125$ (from top to
  bottom, and left to
right).  Note the gradual changes from a
  single
peak (symmetry preserving) to double peaks (symmetry
  breaking) and also
the changes of distortions around each
  peak.}
\label{fig:Hdis}
\end{figure}

\begin{figure}[h!]
\includegraphics[width=8cm]{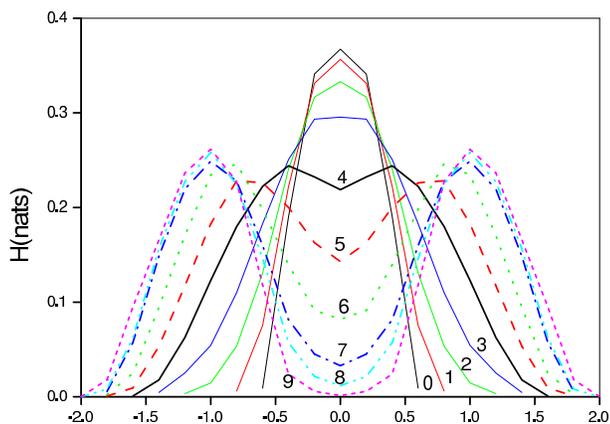}
\caption{Cross sections
  through the Shannon entropy function surfaces for
different
  $\lambda$ values. The curves are labelled with
  $\ell=8\lambda$.
Note that symmetry breaking first occurs when
  $\ell =
4\ (\lambda=0.5)$.}
\label{fig:Hcross}
\end{figure}

Figure \ref{fig:Hdis} shows the distribution of
  the Shannon
entropy function $H(\zeta,\mu)$ for selected values of
  $\lambda$.
When $\lambda$ is small there is a single maximum while
  when
lambda is large enough two maxima are seen. Figure
  \ref{fig:Hcross}
shows cross sections through this $H$ surface for
  all the runs. The
direction of cross section is different for each
  $\lambda$ and is
either chosen to go through the two maxima, or,
  when there is only
a single maximum, it is chosen to go through the
  direction of
minimum curvature. These figures show that there is
  symmetry
breaking in the curve marked 4 ($\lambda=0.5$) and there
  is no
symmetry breaking in the curve marked 3 ($\lambda=0.375$)
  Thus the
critical value of $\lambda$, $\lambda_c$, lies between
  these values, that is somewhat  below $\lambda=0.5$.These
  results confirm that varying the scaling parameter
  $\lambda$
induces symmetry-breaking. 

\begin{figure}[h!]
\includegraphics[width=8cm]{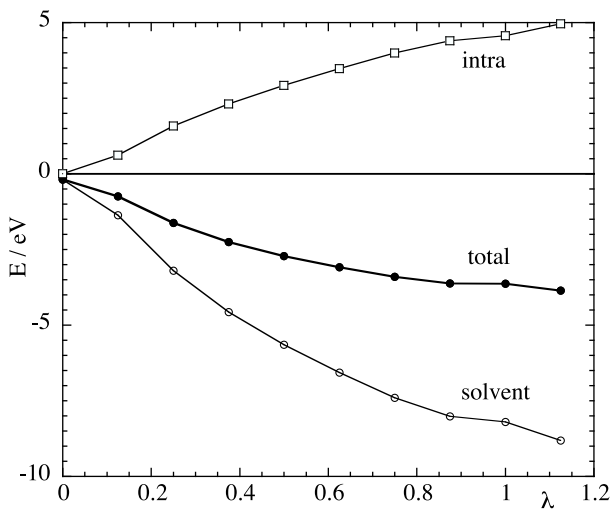}
\caption{Solvent-solute energetics. The upper curve (triangles)
shows
  the change in the average internal energy of the molecule
relative
  to the gas phase which is a measure of the cost of
polarising the
  molecule. The lowest curve (squares) shows the
average
  solute-solvent interaction energy and the middle curve
(circles)
  shows the sum of these two energies, as a function of
$\lambda$.}
\label{fig:en}
\end{figure}

There is no a-priori reason
  to identify $\lambda$ as opposed to some function of $\lambda$ with
  the solvent
strength in the free energy equation
  (\ref{eq:solfe}). However figure \ref{fig:en} shows the
  solvent-triiodide interaction energy (lowest
curve). The
  interaction energy decreases smoothly with the scaling
  factor
$\lambda$ and is approximately linear in the critical
  region. This figure
also shows (upper curve) the energetic cost of
  polarising the ion.

\begin{figure}[h!]
\includegraphics[width=8cm]{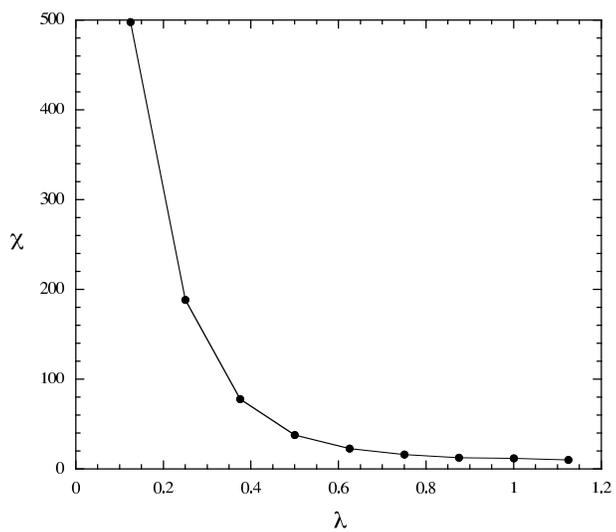}
\caption{Inverse susceptibility of the triiodide ion in solution 
 as a function of
  $\lambda$.}\label{fig:simsus}
\end{figure}

The inverse susceptibilities for the response
  to an external electric
  field
\begin{equation}
\chi^{-1}=kT/<\mu^2>
\end{equation}
is
  plotted in figure \ref{fig:simsus}. It can be seen that  there is a
  monotonic decrease in the inverse susceptibility 
  as a function of the scaling factor $\lambda$, and that $\chi^{-1}$ does not tend to
  zero near the critical value $\lambda
=\lambda_c$. In a
  second-order phase
transition, however, inverse  susceptibilities
  do tend to zero at the
critical point. 

\begin{figure}[h!]
\includegraphics[width=8cm]{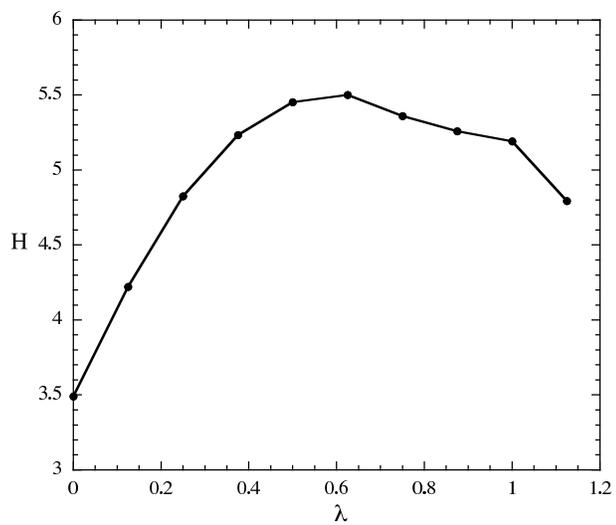}
\caption{Total Shannon entropy $h$ as a function of $\lambda$ for a
  triiodide ion in solution. Note
  that there is a weak maximum at about $\lambda=0.6$ although the
  critical value of $\lambda$ for symmetry breaking lies between
  $\lambda=0.375$ and 0.5. }
\label{fig:htot}
\end{figure}

The Shannon entropy, $h_S$,
  would also be expected to
show critical behavior at the critical
  point. Figure \ref{fig:htot}
shows that in our system this quantity
  has a weak maximum at a value of
$\lambda$ which is slightly
  greater that the critical value $\lambda_c$,
rather than
  diverging at $\lambda_c$.

\section{Discussion}
As the two figures  \ref{fig:htot}, and
 \ref{fig:simsus} do not
show the expected critical behavior
  near the critical value of the
scaling
  constant,
$\lambda=\lambda_C$, we conclude that there is a
  difference between our
system and the mean field Ising model. In
  order to elucidate this
difference  we study the properties of a
  simple model.
 
\subsection{A simple model}
Let us consider a simple model for
  symmetry breaking with a single
order parameter $\eta$ and a Landau
  free energy per molecule given
by
\begin{equation}
\tilde F=a
  (R_0-R) \tilde \eta^2 + C \tilde \eta^4.
\end{equation}
This can
  be rewritten (by  rescaling  $F=C\tilde F/(aR_0)^2$
  and
$\eta^2=C\tilde \eta^2/(2 a R_0)$)
  as
\begin{equation}
F=(1-\rho) \eta^2/2 +
  \eta^4/4,
\end{equation}
where $\rho=R/R_0$  is a measure of the
  relative strength of the
solvent interaction. The critical point
  where symmetry breaking
occurs is $\rho=1$. In order to answer the
  question as to when one
should observe critical phenomena such as
  diverging fluctuations,
we examine the properties of the
  probability distribution of the
order
  parameter
\begin{equation}
p(\eta)=\exp[-\beta F]/Z=\exp[-\beta
  ((1-\rho) \eta^2/2 + \eta^4/4)]/Z 
\label{eq:prob}\end{equation} 
as
  a function of the parameter
$\beta$. In this
  expression
\begin{equation}
Z=\int_{-\infty}^\infty \exp[-\beta
  ((1-\rho) \eta^2/2 +
\eta^4/4)]
  d\eta. \label{eq:z}\end{equation}

\begin{figure}[h!]
\includegraphics[width=8cm]{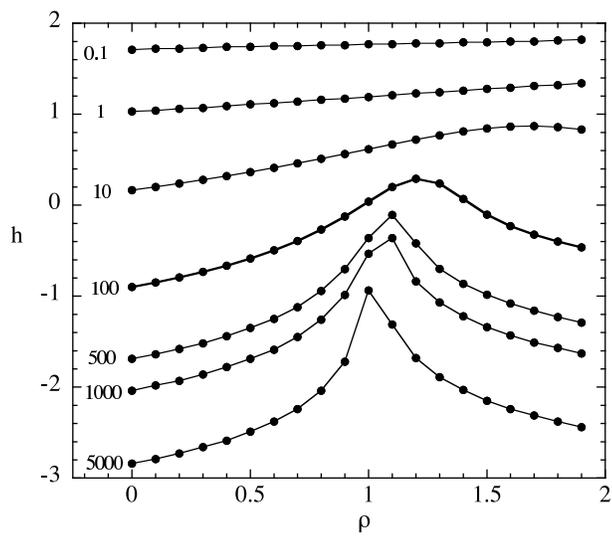}
\caption{Variationof the entropy associated with the order
 parameter,
$h$, 
with relative solvent strength $\rho$ for different
vlaues of the
parameter $\beta$. Note that as $\beta$
increases the maximum gets
sharper and moves closer to
the critical value
$\rho=1$. }\label{fig:modh}
\end{figure}

Figure \ref{fig:modh} shows
  the
values of the  entropy $h=- \int p \ln p \, d \eta$ as
  a
function of $\rho$ for a number of values of $\beta$. It can
  be
seen that the Shannon entropy has a maximum as a function
of
  $\rho$. As $\beta$ increases, this maximum gets sharper and
shifts
  downwards towards $\rho=1$.

\begin{figure}[h!]
\includegraphics[width=8cm]{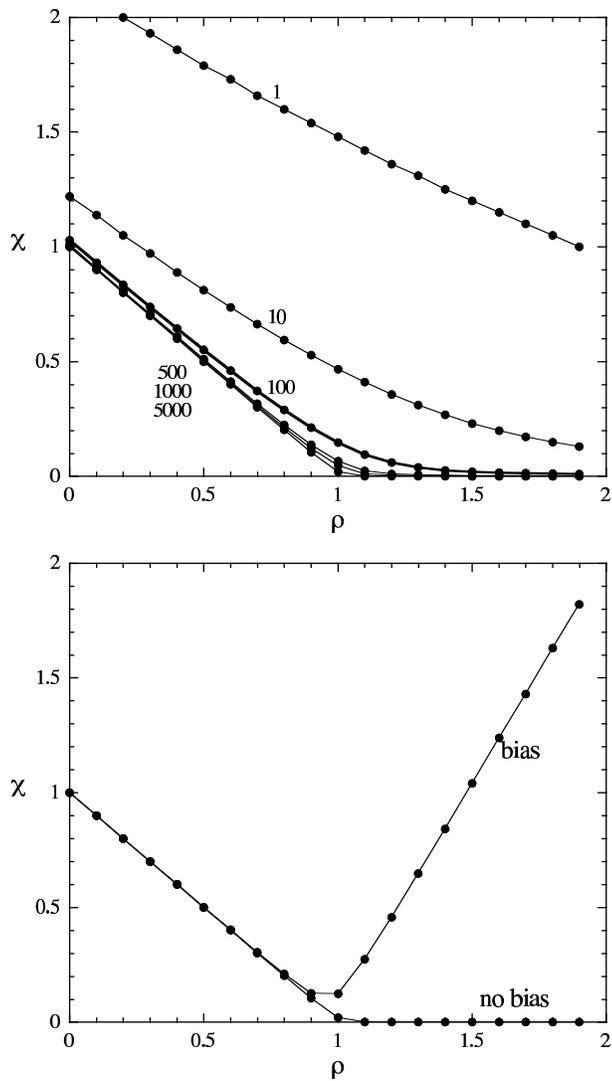}
\caption{Inverse
  susceptibility of the model as a
function of the solvent
  strength. Above: reults for various values of the
paperameter
  $\beta$.  Below: Results for a high value
of $\beta$ with and
  without a small biassing field.
Note that with the biassing field
  the model has a
minimum inverse susceptibilty near the
critical
  value $\rho=1$. In the absence of the
biassing field there is a
  linear decrease in the inverse
susceptibilty between $\rho=0$ and
  $\rho=1$. }\label{fig:sus}
\end{figure}

For a second order phase transition, the inverse of
  the
susceptibility $\chi$, which is given by $\chi^{-1} =
  1/\beta \langle \eta^2 \rangle $, tends to zero at the critical
  point
where the phase transition occurs. The upper part of
  figure
\ref{fig:sus} shows the values of the inverse of
  the
susceptibility of the model system as a function of $\rho$ for
  a
range of values of the parameter $\beta$. Again we observe that
  if
$\beta$ is large enough the inverse susceptibility decreases
  linearly
towards a value of zero at $\rho < 1$. A simple
  calculation of the
susceptibility above the critical value of
  $\rho$ is misleading as the
probabilities of positive and negative
  values of the order parameter are
equal and the mean value is
  always 0. However if a small biassing field
$b$ is added giving the
  free energy
\begin{equation}
F=-b \eta + (1-\rho) \eta^2/2 +
  \eta^4/4,
\end{equation}
then the lower part of figure
  \ref{fig:sus} shows that the inverse
susceptibility  increases with
  $\rho$ above the critical
value $\rho=1$.

The results at large
  $\beta$ are very similar to those obtained
from a mean field model
  of a second order phase transition with
Landau free
  energy
\begin{equation}
F= -b \eta + a(T_0-T) \eta^2/2 + C
  \eta^4/4.
\label{eq:fept}\end{equation} In the standard treatment
  of this model \cite{landau}, one is only concerned with the minima of the
  function. One
finds that above the critical temperature $T_0$ the
  minimum of $F$
is at $\eta_{min}=0$ in the absence of an external
  field, while
below $T_0$ the minima are at $\eta_{min} =
  \pm
(a(T_0-T)/C)^{1/2}$. The susceptibility is calculated from
  the
change of the position of $\eta_{min}$ with the strength of
  the
external field, $\chi = d \eta_{min}/d b$. This
  gives
\begin{eqnarray}
\chi^{-1}=a(T-T_0)         & & T > T_0
  \nonumber \\
\chi^{-1}=2 a (T_0-T)      & & T <
  T_0.
\end{eqnarray}
It will be noted that the standard treatment
  does not include
fluctuations in the order parameter, while the
  expression that we
have used (equation \ref{eq:prob}) gives the
  full range of
possible values. The reason that one can ignore
  fluctuations in
the treatment of phase transitions is that the
  expression for the
free energy given in equation (\ref{eq:fept}) is
  the free energy per
unit cell or per spin. The observed $\eta$ is
  the average over all
the unit cells or all the spins and, as one
  observes the average
over a large number $N$ of unit cells or over
  $N$ spins, the
probability of observing a given value of $\eta$ is
  given by
\begin{equation}
p(\eta)=\exp[-(N/kT) (a(T_0-T) \eta^2/2
  + C\eta^4/4)]/Z.
\label{eq:ptprob}\end{equation} In the limit $N
  \rightarrow
\infty$ it is indeed only the minima that are
  observed. In the
solvation-induced symmetry breaking situation we
  observe
individual molecules and the relevant probability is given
  by the
similar equation \ref{eq:prob} but with the difference
  that
$\beta=(4a R_0)^2/(CkT)$ rather than $N/kT$. The fact that we
  only
see significant maxima  in the Shannon
entropy near the
  critical value of $\rho$ when $\beta$ is large is
consistent with
  the fact that true critical phenomena only occur
in the limit of $N
  \rightarrow \infty$ and depend on the
observation of an order
  parameter which is an average over many
replicated systems. While
  solvent-induced symmetry breaking will
never show true divergences,
  there will be maxima in the Shannon
entropy if the value of the
  parameter $(aR_0)^2/C$ is large enough
compared to $kT$. Comparing
  the model results with those observed for the
triiodide ion, we
  estimate that the order of magnitude of $(aR_0)^2/CkT$
is about
  100. At this value the is a weak maximum in the entropy at a
higher
  solvent strength than the critical one and the
  inverse
susceptibility decreases smoothly through the critical
  point.

\section{Conclusion}
In this paper we have examined the
  solvent-induced symmetry
breaking induced by water and modified
  water. The use of scaled
charges in the solvent models allows us to
  vary the solvent
strength which induces the symmetry breaking
  continuously and to
determine whether there are any phenomena
  analogous to critical
phenomena. In this particular system the
  transition from no
symmetry breaking to symmetry breaking is weak
  and
there is only a small maximum in the Shannon entropy. By
  comparing the
results to a simple model we see that if
  $\beta=(aR_0)^2/CkT\approx 100$
the model system shows a rather
  similar behavior.
In order to see significant critical behavior
  this parameter would need to be
larger by a factor of 10 or
  more.

Although this example of solvent-induced symmetry breaking
  does not show
critical effects, there may be other situations which
  do. A system
which was less polarisable, would have a larger value
  of $aR_0$, but
would need a strong interaction with the solvent for
  symmetry-breaking to
occur at all. 

\section*{Acknowledgments}
 We thank EPSRC for financial support (grant GR/N38459/01).

\section*{Appendix}
Standard methods of statistical mechanics shows
that in contact with a
heat bath the probability $p({\bf r}) d{\bf
r}$ of observing the $3N$
system coordinates ${\bf r} = \{r_i\}$
with values  between  ${\bf r}$
and  ${\bf r + dr}$ is given
by
\begin{equation}
p({\bf r}) = \exp[-\beta H({\bf
r})]/Z,
\label{eq:pconf}\end{equation}
where $H$  is the
potential energy.  $Z$ is the configurational
integral  for the
complete system
\begin{equation}
Z =\int \exp[-\beta H({\bf
r})]\,d{\bf r}  = \exp[-\beta F_{\rm
conf}],
\end{equation}
which defines $F_{\rm conf}$, the
classical configurational part of the
Helmholtz free energy of
the
system.

The total Helmholtz free energy can be
written
\begin{equation}
F= -kT \ln Z -kT \ln A = F_{\rm conf}+
 F_{mom}
\end{equation}
where $F_{mom}=-kT\ln A$ is  an additional
ideal term due to
the momentum and
indistinguishability,
\begin{equation}
A=\prod_i ( 2 \pi m_i kT/
 h^2_i)^{3/2}/N_i!
\end{equation}
where the product is over
different types of atom, $i$.

Let us define the  Landau free
energy $F_L(\eta)$ as the free energy of
the system when an order
parameters $\eta ({\bf r})$ is constrained to a
fixed value
$\eta_0$. Thus
\begin{equation}
F_L(\eta_0)= -kT \ln \left[\int
 \exp[-\beta H]
 \delta(\eta-\eta_0)\,d{\bf
r}
\right].
\end{equation}

From
equation (\ref{eq:pconf}) we see that the probability $p(\eta)
d\eta$
of observing the value of the order parameter between $\eta$
and
$\eta+ d\eta$ in the unconstrained system is
\begin{equation}
p(\eta)\,d\eta= \exp[-\beta
 F_L(\eta)]\,d\eta/Z
\end{equation}
with
\begin{equation}
Z=\int
 \exp[-\beta F_L(\eta)]  d\eta.
\end{equation}

Now let us
consider the quantity
\begin{equation}
h = -\int p(\eta) \ln
p(\eta)\, d\eta.
\end{equation}
$h$ is obviously a property of
the system (molecule plus bath) rather
a property of our
knowledge about it. In the words of Denbigh
and
Denbigh\cite{denbigh} it is an objective property rather than
a subjective
property. As
$p(\eta)$ is a density rather than a
dimensionless property, $Z$ has the
dimensions of the order
parameter and  so the zero of $h$ depends on
the units
of
$\eta$.

Using the expressions
above

\begin{equation}
h  =  \beta \left [\int p(\eta)
F_L(\eta)d\eta - F_{\rm
conf}\right]
\end{equation}
or
\begin{equation}
h =  \beta
[\langle F_L(\eta)\rangle - F_{\rm conf}],
\end{equation}
the
difference of the average Landau free energy and
the
configurational free energy of the unconstrained
system. However as
$F=U-TS$ and the average energy is the same
whether calculated by
averaging over the constrained systems or
over the unconstrained system,
$h$ is the difference of the
entropy of a  constrained system
averaged over all constrained
systems and the total entropy,
  
\begin{equation}
h =   [\langle S_{L}(\eta)\rangle - S_{\rm
 conf}]/k_B.
\end{equation}
Here $S_{L}$ is the entropy of a
 constrained system with
fixed $\eta$, and the
angular brackets
 denote an average over all constrained systems
weighted with their
 probabilities
in the unconstrained system. Thus $h$ can be
 interpreted as the
entropy associated with the order parameter
 distribution.

The Shannon information entropy is defined in
 terms of the information
that one has about the system. For example
 if a histogram of
probabilities of $\eta$ is constructed with bin
 widths $\delta$ and
if one calculates the Shannon entropy using
 natural logarithms rather
than logarithms to base
 $2$,
\begin{equation}
h_S = -\sum p_i \ln
 p_i,
\end{equation}
then
\begin{eqnarray}
h_S &\approx& \int p
 \ln p \,d\eta - \ln (\delta )\nonumber \\
&=&  h - \ln (\delta
 ).
\end{eqnarray}
 
Thus the Shannon entropy, $h_S$, which is a property of the
 information
collected about the system,  provides an approximation
 to the system
property $h$. The latter is a measure of the spread of
 the order
parameter. This treatment may readily be extended to more
 than one order
parameter.

\suppressfloats \pagebreak
%\end{document}
\begin{table}
\caption{Intermolecular Lennard-Jones site-site parameters
  for
water. Cross terms in the Lennard-Jones interactions
  were
calculated using the Lorentz-Bertholet rules.}
\label{tab:params}
\begin{center}
\begin{tabular}{ccccc}
\hline atoms $i$ &  $\epsilon_{ii}$/kJ\,mol$^{-1}$
&
$\sigma_{ii}$/\AA   &  q($i$)/e  \\ \hline
O         &  0.6502                           &   3.169             &
$-0.8476\lambda$  \\
H         &  0                                &   0                 & 

$0.4238\lambda$  \\
I         &  0.4184                           &   5.167             &
varies  \\
\hline
\end{tabular}
\end{center}
\end{table}

\end{document}